\documentclass[11pt]{article}
\usepackage{amsmath,amssymb}
\usepackage{array}
\numberwithin{equation}{section}

\begin{document}
\baselineskip 100pt
\renewcommand{\arraystretch}{0.666666666}
\large
\parskip.2in


\newcommand{\beq}{\begin{equation}}
\newcommand{\eeq}{\end{equation}}
\newcommand{\eqalinb}{\begin{eqnarray}}
\newcommand{\eqaline}{\end{eqnarray}}
\numberwithin{equation}{section}

\newcommand{\dg}{\dagger}
\newcommand{\acc}{\\[3mm]}
\newcommand{\Ref}[1]{(\ref{#1})}
\def\mod#1{ \vert #1 \vert }
\def\chapter#1{\hbox{Introduction.}}

\def\exp{\hbox{exp}}
\def\Ln{\hbox{ln}}
\def\arctgh{\hbox{arc\,tanh}}


\def\a{\alpha}
\def\b{\beta}
\def\g{\gamma}
\def\d{\delta}
\def\ep{\epsilon}
\def\e{\varepsilon}
\def\z{\zeta}
\def\t{\theta}
\def\k{\kappa}
\def\l{\lambda}
\def\s{\sigma}
\def\f{\varphi}
\def\w{\omega}
\def\vx{{\textbf{\hbox{x}}}}
\def\vv{{\textbf{\hbox{v}}}}
\def\vJ{{\textbf{\hbox{J}}}}
\def\vH{{\textbf{\hbox{B}}}}
\def\Fm{\boldsymbol{\rm{F}_m}}
\def\vw{\boldsymbol{\w}}
\def\v{{\hbox{v}}}
\def\J{{\hbox{J}}}
\def\H{{\hbox{B}}}
\def\u{{\hbox{u}}}
\def\x{{\hbox{x}}}
\def\ve{{\textbf{\hbox{e}}}}
\def\div{\nabla\!\cdot}
\def\rot{\nabla\!\times}
\def\grad{\nabla}
%
\def\Dp#1#2{\frac{\partial #1}{\partial #2}}
\def\dr#1{\frac{\hbox{d}#1}{\hbox{d} r}}
\def\dt#1{\frac{\hbox{d}#1}{\hbox{d} t}}
\def\ds#1{\frac{\hbox{d}#1}{\hbox{d} s}}
\def\Dt#1{\frac{\partial #1}{\partial t}}
\def\dss#1{\frac{\hbox{d}}{\hbox{d}s}\!\left[#1\right]}
\def\df{\hbox{d}}
%
%
%
\def\dm{\frac{1}{2}}
\def\N{\mathbb{N}}
\def\Z{\mathbb{Z}}
\def\R{\mathbb{R}}
\def\de{\delta|\vec\l|}
\def\dei#1{\delta_{#1}\,|\vec\l|}
\def\ol#1{{\overline{#1}}}
\def\dv#1{\>\>d {#1}}
\def\dl{\vec {d\ell}}
\def\ldts{{\ldotp\ldotp\ldotp}}
\def\ld{\,\ldots\,}
\def\ddots#1{\buildrel\ldts\over{#1}}
\def\/#1#2{\frac{#1}{#2}}
\def\vec#1{{\boldsymbol{#1}}}

%
%
%
\def\[{\left [}
\def\]{\right ]}
\def\qd{$\quad$}
\newcommand{\ie}{{i.e.,}}
\def\cf{\hbox{c.f.,}{}}
\def\eg{\hbox{e.g.,}{}}


\title{\bf Reduction and Exact Solutions of the Ideal Magnetohydrodynamic
Equations}
\author{
P. Y. Picard\thanks{email address: picardp@inbox.as}\\
D\'epartement de Physique, Universit\'e de Montr\'eal,\\
C. P. 6128, Succ.\ Centre-ville, Montr\'eal, (QC) H3C 3J7, Canada}
\date{}

\maketitle
\begin{abstract}
In this paper we use the symmetry reduction method to obtain
invariant solutions of the ideal magnetohydrodynamic equations in
(3+1) dimensions. These equations are invariant under a
Galilean-similitude Lie algebra for which the classification by
conjugacy classes of r-dimensional subalgebras ($1\leq r\leq 4$) was
already known. So we restrict our study to the three-dimensional
Galilean-similitude subalgebras that give systems composed of
ordinary differential equations. We present here several examples of
these solutions. Some of these exact solutions show interesting
physical interpretations.

\vskip0.4cm


\end{abstract}

\section{Introduction}
The ideal magnetohydrodynamic (MHD) equations are the most basic
single-fluid model with the Maxwell's equations for describing the
interactions between magnetic and pressure forces within an
electrically conducting fluid for long spatial scale and low
frequency phenomena in plasma physics. There are not many exact
solutions in the theory of ideal magnetized fluid governed by the
system of MHD equations. We present here some exact analytic
solutions of this system. By exact solution we mean one obtained by
a reduction of the full set of partial differential equations (PDEs)
of the original system to one or more ordinary differential
equations (ODEs) involving fewer independent variables. This can be
achieved by applying the symmetry reduction method (SRM) based on
group theory. This method for reducing the number of independent
variables in an equation is to require that a solution be invariant
under some subgroups of the Lie symmetry group of the given system;
the so-called G-invariant solution (GIS).

The paper is organized as follows. Section 2 briefly gives a general
description of the ideal MHD equations. In Section 3 contains a
short summary of the SRM and its application to the MHD system. In
Sections 4 and 5, some examples of GIS are presented and also
interpreted by a physical point of view for two configurations of
the magnetic field: $\vH=({\H}_1,{\H}_2,0)$ and
$\vH=({\H}_1,{\H}_2,{\H}_3)$, respectively. Section 6 summarizes the
obtained results which suggest further applications.

\section{Preliminary information}
The fluid under consideration is assumed to be ideal and perfectly
conductive, that is, neglecting all the dissipative and dispersive
effects, such as viscosity, magnetic resistivity, thermal
conductivity and Hall effect. The fluid is assumed to be unbounded
and the medium is non magnetic, so its magnetic permeability $\mu$
is taken to be one. Furthermore we suppose that the fluid is
described by a perfect gas equation of state. Under the above
assumptions the ideal MHD model is governed by the system of nine
PDEs:\eqalinb
\/{\df \rho}{\df t}+(\div\vv)\rho=&0\,,\label{s1}\\
\/{\df\vv}{\df t}+\rho^{-1}\,\grad p+
\rho^{-1}\,(\vH\times\vJ)=&\boldsymbol{0}\,,\label{s2}\\
\/{\df p}{\df t}+\g(\div\vv)p=&0\,,\label{s3}\\
\Dt{\vH}-\rot\big(\vv\times\!\vH\big)=&\boldsymbol{0}\,,\label{s4}\\
\div\vH=&0\,,\label{s5} \eqaline where $\df/\df t=\partial/\partial
t+(\vv\cdot\grad)$ is the convective derivative. Here $\rho$ is the
mass density, $p$ is the pressure, $\vv=\big(\v_1,\v_2,\v_3\big)$
and $\vH=\big({\H}_1,{\H}_2,{\H}_3\big)$ are the flow velocity and
the magnetic field, respectively and $\g$ is the adiabatic exponent.
Note that throughout this paper we denoted by $\ve_1$, $\ve_2$ and
$\ve_3$ the unit vectors in the direction of the $x$-, $y$- and
$z$-axes, respectively. The current density is given by the
Amp\`ere's law \beq\vJ=\rot\vH\,,\eeq for which the displacement
current is negligible since the flow is non-relativistic. All the
functions $\u=\big(\rho,p,\vv,\vH\big)$ depend on time $t$ and
coordinates $\vx=(x,y,z)$. The system of MHD equations
\Ref{s1}--\Ref{s5} is quasilinear, hyperbolic, and is written in the
normal (Cauchy-Kowalewski) form. In equation \Ref{s2}, there is no
extraneous force other than the Lorentz force (\ie\,the magnetic
force)\beq \Fm=\vJ\times\vH\,,\label{lorentz}\eeq which can be
broken down into a magnetic pressure force and a magnetic tension
force \beq\Fm=-\/{1}{2}\grad\big(|\vH|^2\big)
+(\vH\cdot\nabla)\vH\,.\label{em}\eeq It is noteworthy to recall
that the integral form of Faraday's law \Ref{s4} \beq\frac{\df }{\df
t}\left(\int_{{\cal
S}}{\vH}\cdot{\df\vec{\rm{s}}}\right)=0\,,\label{th.f}\eeq where
${\cal S}$ is an arbitrary surface moving with the fluid,
corresponds to the magnetic flux conservation law, \ie\, the
``Alfv\'en's frozen-in'' theorem \cite{alf1}, and has some
fundamental consequences. The magnetic field lines are ``glued'' to
the medium, and transported entirely by convection and there is no
diffusion of $\vH$ through the conducting media. So, in
``flux-freezing'' regime, the convective derivative of $\vH/\rho$
can be written as \cite{landau}: \beq\/{\df}{\df
t}\left(\/{\vH}{\rho}\right)-
\left(\/{\vH}{\rho}\cdot\grad\right)\vv=\boldsymbol{0}\label{lines}\,.\eeq
This says that $\vH/\rho$ evolves in the same manner as the
separation $\boldsymbol{\delta l}$ between two points in the fluid.
Notice that the vorticity $\vw$ of the flow and $\vH$ are physically
related by:\beq\Dt{\vw}-\rot\big(\vv\times\!\vw\big)
=\rot\left[\/{\Fm}{\rho}\right]\label{vort}\,.\eeq In hydrodynamics
where $\vH=\boldsymbol{0}$, if at one instance $\vw=\boldsymbol{0}$,
the flow will be irrotational for all subsequent times. So, the
vorticity field obeys a transport equation similar to \Ref{s4}: the
vortex lines are frozen into the fluid matter. However this result
does not hold for ideal MHD because the term
$\rot\left[{\Fm}/{\rho}\right]$ is in general not equal to zero, and
may induce vorticity even though the flow is initially irrotational.

\section{The symmetry of the magnetohydrodynamic equations}
The language of group theoretical methods for studying systems of
PDEs is a very useful and suitable tool for investigating the main
features of numerous problems in various branches of mathematical
physics. The task of finding an increasing number of solutions of
systems of PDEs is related to the group properties of these
differential equations. Its main advantages appear when group
analysis makes it possible to construct regular algorithms for
finding certain classes of solutions without referring to additional
considerations but proceeding only from the given systems of PDEs. A
systematic computational method for constructing the group of
symmetries for given system of PDEs has been extensively developed
by many authors \cite{ovs}--\cite{clark}. A broad review of recent
developments in this subject can be found in several books as
\cite{ibr}, \cite{clark}. The methodological approach adopted here
is based on the symmetry reduction method for PDEs invariant under a
Lie group ${\cal G}$ of point transformations. This means that the
groups under considerations are the connected local transformation
groups acting on the space of $p$ independent and $q$ dependent
variables $E\times U$,\;\ie\beq \widetilde\x=\Lambda_{{\cal
G}}\,(\x,\u),\qquad \widetilde\u=\Omega_{{\cal G}}\,(\x,\u)
\;;\qquad\x=(x^1,\ld,x^p),\;\quad\;\u=(u^1,\ld,u^q)\,.\eeq By a
symmetry group of a system $\Delta$ of differential equations we
understand a local Lie group ${\cal G}$ transforming both the
independent and dependent variables of the considered system of
equations in such a way that ${\cal G}$ transforms given solutions
$\u\,(\x)$ of $\Delta$ to the new ones $\widetilde\u(\widetilde\x)$
of $\Delta$. The Lie algebra ${\cal L}$ of ${\cal G}$ is realized by
the vector fields $\hat X$ which for the system composed of $m$ PDEs
of order $k$ \beq\Delta^l\,\big(\x,\u^{(k)}\big)=0,\qquad
l=1,\ld,m\eeq where $\u^{(k)}$ denotes all partial derivatives of
$\u$ up to order $k$, can be expressed as follows \beq\hat
X=\xi^\mu\,(\x,\u)\partial_{x^\mu}+\phi^j\,(\x,\u)\partial_{u^j}\,,
\eeq where $\xi^\mu$ and $\phi^j$ are assumed to be functions of
$(\x,\u)\in E\times U$ only. The functions $\xi^\mu$ and $\phi^j$
are defined by the infinitesimal invariance criterion
\cite{olv}\beq{\rm pr}^{(k)}\big(\hat
X\big)\Delta^l\Big|_{\Delta^{n}=0}=0,\qquad l,n=1,\ld,m\eeq where
${\rm pr}^{(k)}(\hat X)$ is the $k$ prolongation of the vector
field. There exist standard algorithms for determining the symmetry
algebra ${\cal L}$ and classifying subalgebras ${\cal L}_p$ of
${\cal L}$ \cite{clark}.

The symmetry algebra of the MHD equations \Ref{s1}--\Ref{s5},
denoted by ${\cal M}$, has been found by Fuchs \cite{fuchs}, and
independently by Grundland and Lalague \cite{amg0}. It is spanned by
the following 13 Galilean-similitude (GS) infinitesimal generators:
\begin{align}
P_{\mu}&=\partial_{x_\mu}\,,\qquad
J_{k}=\epsilon_{kij}\big(x_i\partial_{x_j}+\v_{j}\partial_{\v_{j}}+
{\H}_{i}\partial_{{\H}_j}\big)\,,\qquad
K_{i}=t\partial_{x_i}+\partial_{\v_i}\,,\label{gene}\\
F&=t\partial_{t}+x_{i}\partial_{x_i}\,,\quad
G=-t\partial_{t}-2\rho\partial_{\rho}+\v_i\partial_{\v_i}\,,\quad
H=2\rho\partial_{\rho}+2p\partial_{p}+{\H}_{i}\partial_{{\H}_i}\,,\nonumber
\end{align} where $\epsilon_{kij}$ is the Levi-Civita
symbol, $i,j,k=1,2,3$; $\mu=0,1,2,3$. Thus, the MHD equations
\Ref{s1}--\Ref{s5} are invariant under time ($P_o$) and spatial
($P_i$) translations, rotations ($J_k$), Galilei transformations
($K_i$), dilations ($F$, $G$, and $H$ which is the center of ${\cal
M}$; $H$ commutes with all the generators). The basis of ${\cal M}$
and the commutator tables are given in \cite{amg0}. In contrast to
the results obtained for the $(1+1)$ and $(2+1)$ dimensional
versions of the MHD model \cite{fuchs}, the dimension of the Lie
algebra ${\cal M}$ for MHD equations \Ref{s1}--\Ref{s5} in the full
$(3+1)$ dimensions ($x$, $y$, $z$ and time $t$) is independent of
the value of the adiabatic exponent $\g$ that we treat here as a
fixed parameter: $\g\geq 1$.

A classification by conjugacy classes of $r$-dimensional subalgebras
($1\leq r\leq 4$) corresponding to the GS algebra ${\cal M}$ has
been established by Grundland and Lalague \cite{amg0}. Their
investigation is limited to subgroups of dimension not greater than
the number of independent variables. This constraint is imposed by
the SRM itself, in order to reduce the initial system of PDEs to a
lower dimensional system of PDEs which now involves $p-r$
independent variables. The method of classification which they have
used was developed by Patera \textit{et al} \cite{pater}, and
published in a series of papers in the late 1970's. More recently,
this subject has been treated in such books as \cite{ibr},
\cite{clark}.

The aim of this work \cite{these} was to fulfill the last steps of
the SRM applied to the MHD equations \Ref{s1}--\Ref{s5}. Obtaining
the invariant solutions, called G-invariant solutions (GIS), under
some subgroup ${\cal G}_{p}\subset {\cal G}$ requires the knowledge
of the group invariant ${\cal I}$ of ${\cal G}_p$. For each of the
104 three-dimensional subalgebras of ${\cal M}$ taken from
\cite{amg0}, we have calculated a set of functionally independent
invariants. Suppose $\{v_1,v_2,v_3\}$ is a basis of infinitesimal
generators of the Lie algebra ${\cal L}_3$, then ${\cal I}$ is a
${\cal L}_3$-invariant if and only if $v_i({\cal I})=0$, $i=1,2,3$.
Thus, for each algebra, we obtain nine functionally invariants. We
can calculate invariant solutions if the invariants are of the form
\beq{\cal I}=\big\{s(\x)\,;\,I_i\,(\x,\u)\big\},\qquad
i=1,\ld,8\label{step2}\eeq such that \beq{\rm{rank
}}\left(\Dp{I_i\,(\x,\u)}{\u}\right)=8\,.\label{step3}\eeq In our
investigation, we find that the invariants of twelve
three-dimensional algebras among those classified by Grundland and
Lalague \cite{amg0} do not have the last properties. The solutions
corresponding to these algebras are called partially invariant
solutions (PIS) and therefore the SRM is inapplicable. So, we
concentrate here only on invariant solutions.

From the expression given in \Ref{step2}, we form the equation \beq
I_i\,(\x,\u)={\cal F}_i\,\big(s(\x)\big)\,,\qquad
i=1,\ld,8\label{step4}\eeq where ${\cal F}_i$ are arbitrary
functions, giving the orbits of an invariant solution. From
\Ref{step3} and the implicit functions theorem, we obtain \beq
\u=\phi_i\big({\cal F}_j\,(s),\x\big)\,,\qquad
i,j=1,\ld,8\label{step5}\eeq where $\phi_i$ are arbitrary functions.
Substituting each of the expressions \Ref{step5} into equations
\Ref{s1}--\Ref{s5}, we can reduce the MHD system to a new system
composed of ODEs. Such system is called the reduced system. The
variable $s$ is the symmetry variable and plays the role of the
independent variable of the reduced system. Solving the reduced
system and inserting these solutions into \Ref{step5}, we obtain
finally the GIS of the system \Ref{s1}--\Ref{s5}.

In the two next sections, we present ten examples of GIS that we
have obtained from the GS subalgebras of dimension three \cite{amg0}
($G_i$ refers to the solution corresponding to this algebra given in
the title of each subsection). These solutions are also classified
by some of their physical characteristics and the types of the waves
that they may generate.

\section{$\boldsymbol{\vH=\big({\H}_1,{\H}_2,0\big)}$}\label{sec1}
\subsection*{$\boldsymbol{G_1=\big\{J_3+K_3+\a H,\,P_1,\,P_2\big\}}$,
\;\quad\;$\a\in\R$}\label{3-15} From this algebra, we have obtained
the corresponding G-invariant solution
\begin{align}
\rho&=\exp\[2\a\/{z}{t}\]R(t),\qquad p=A_o\,
t^{(1-\g)}\rho\,,\label{3.15}\\
\v_1&=U_o\sin\left(\/{1}{2\a}\ln\[\/{\theta_o}{tR(t)}\]-\/{z}{t}\right),\;\;
\v_2=U_o\cos\left(\/{1}{2\a}\ln\[\/{\theta_o}{tR(t)}\]-\/{z}{t}\right),\;\;
\v_3=\displaystyle\/{z}{t}-\/{W(t)}{t}\,,\cr
{\H}_1&=X_o\sqrt{\/{\rho}{t}}
\sin\left(\/{1}{2\a}\ln\[\/{\theta_o}{tR(t)}\]-\/{z}{t}\right)\!,\;\;
{\H}_2=X_o\sqrt{\/{\rho}{t}}
\cos\left(\/{1}{2\a}\ln\[\/{\theta_o}{tR(t)}\]-\/{z}{t}\right)\!,\;\;
{\H}_3=0\,,\nonumber\end{align} where $U_o$, $X_o$ and
$\theta_o\in\R$. Note that throughout this paper, we denoted the
arbitrary constants by $(\;)_o$. The functions $R$ and $W$ are given
by:
\begin{align} R(t)&=\left\{\begin{array}{ll}
\displaystyle {R_o}\,{t^{(4\a^2
A_o-1)}}\,\exp\[-\/{2\a}{t}W_o-\/{2\a^2}{t}{X_o^2}\big(1+\ln
[t]\big)
\]\;\;\hbox{for}\;\;\g=1\,,\\
\displaystyle\/{R_o}{t}\exp\[-\/{2\a}{t} W_o-\/{2\a^2}{t}\big(2
A_o+X_o^2\big)\big(1+\ln [t]\big)\]\;\;\hbox{for}\;\;\g=2\,,\nonumber\\
\displaystyle\/{R_o}{
t}\,\exp\[-\/{2\a}{t}W_o-\/{2\a^2}{t}{X_o^2}\big(1+\ln
[t]\big)+\displaystyle\/{4\a^2
A_o}{(2-\g)(1-\g)}t^{(1-\g)}\]\;\hbox{for}\;\g\neq 1,2\,,
\end{array}\right.\\
W(t)&=\left\{\begin{array}{ll} \displaystyle
W_o+\a\big(2A_o+X_o^2\big)\ln
[t]\;\;\hbox{for}\;\;\g=2\,,\\
\displaystyle W_o+\/{2\a A_o}{(2-\g)}t^{(2-\g)}+\displaystyle\a
X_o^2\ln[t]\;\;\hbox{for}\;\;\g\neq 2\,,\nonumber
\end{array}\right.\end{align}
with $R_o\in\R^{+}$, $\a\in\R/\{0\}$ ; $0\leq A_o<1/4\a^2$ for
$\g=1$ and $A_o\in\R^{+}$ for $\g>1$, $W_o\in\R$ with
$\rm{sgn}[W_o]=\rm{sgn}[\a]$. Under these conditions and for $t>0$,
the solution $G_1$ is non-singular and tends to zero for large
values of $t$.

This solution can be interpreted physically as a nonstationary and
compressible fluid for which the shape of the flow is a helix of
radius $U_o$ and pitch angle $\phi=\arctan\big[\v_3/U_o\big]$.
Indeed, the motion described by the solution $G_1$ is a circular
motion in the $x$-$y$ plane (where lies the magnetic field $\vH$)
and an accelerated motion parallel to the $z$-axis resulting from
the action of the Lorentz force: \beq
\Fm=-\/{1}{2}\grad\big(|\vH|^2\big)=-\a{X^2_o}\/{\rho}{t^2}\,\ve_3\,.\label{grad}\eeq
The Lorentz force $\Fm$ acts perpendicularly to $\vH$, causing
compressions and expansions of the distance between lines of force
without changing their direction, as does a magnetoacoustic fast
wave ($\rm{F}$) which propagates perpendicularly to $\vH$ in an
ideal fluid \cite{pic}. There is no tension force acting on the
magnetic fields lines. So, the classical Bernoulli theorem is still
valid with the pressure of the fluid is replaced by $p+(|\vH|^2/2)$
which is then the total pressure. The solution $G_1$ has vorticity
that lies in the $x$-$y$ plane, and referring to \Ref{vort}, the
vortex lines move with the fluid. Consequently, by virtue of
Kelvin's theorem the circulation \beq\Gamma_{{\rm c}}=\oint_C
{\vv}\cdot{\df\vec{l}}\label{circ}\eeq around a fluid element is
preserved \cite{batch}. Also, the solution $G_1$ is characterized by
the fact that $(\vH\cdot\nabla)\vv=\boldsymbol{0}$, which shows that
the velocity of the fluid $\vv$ is constant along each line of force
but also meaning that $\vH/\rho$ is conserved along the flow, and
consequently the magnetic field lines remains inextensible
[\cf\,\Ref{lines}]. The current density induced by the magnetic
field $\vH$ is equal to \beq \vJ=-\/{1}{t}\big[\a\H_2+
\H_1\big]\ve_1+\/{1}{t}\big[\a\H_1-\H_2\big]\ve_2\,,\eeq which lies
in the $x$-$y$ plane and always has a nonzero component.
\subsection*{
$\boldsymbol{G_2=\big\{J_3+K_3+\a_1 P_3+\a_2 H,\,
K_1,\,K_2\big\}}$,\;\quad\;$\a_1$, $\a_2\in\R$}\label{3-9} With
respect to this algebra, we have calculated the corresponding
invariant solution
\begin{align}
\rho&=\/{R_o}{t^2(t+\a_1)}\exp\[R(t)-\/{[5\a_1+W_o-2\a_2
z]}{(t+\a_1)}
-\/{2\a_2^2{X^2_o}}{(t+\a_1)}\big(1+\ln[t+\a_1]\big)\]\,,\cr
p&=A_o\big[t^2(t+\a_1)\big]^{(1-\g)}\rho\,,\nonumber\\
\v_1&=\/{x}{t}-\/{U_o}{t}\sin\theta(z,t),\qquad\;\;
\v_2=\/{y}{t}-\/{U_o}{t}\cos\theta(z,t)\,,\nonumber\\
\v_3&=\/{z}{(t+\a_1)}-\/{2\a_2}{(t+\a_1)}
W(t)-\/{A_o\a_2{X^2_o}}{(t+\a_1)}
{\ln[t+\a_1]}-\/{(5\a_1+W_o)}{2\a_2(t+\a_1)}\,,\label{3.9}\\
{\H}_1&=\/{X_o\sqrt{\rho}}{\sqrt{t+\a_1}}\sin\theta(z,t),\qquad
{\H}_2=\/{X_o\sqrt{\rho}}{\sqrt {t+\a_1}}\cos\theta(z,t),\qquad
{\H}_3=0\,,\nonumber
\end{align}
where $R_o,$ $A_o\in\R^{+}$; $W_o$, $U_o$ and $X_o\in\R$. The
expressions for the functions $R$ and $W$ depend on specific values
of $\g$.
\par
\noindent For $\g=3/2$:\begin{align}
R(t)=&\left(\/{\sqrt{t+\a_1}-\sqrt{\a_1}}{\sqrt{t+\a_1}+\sqrt{\a_1}}
\right)^{2\a_2^2A_o/(\a_1)^{{3}/{2}}} \exp\!\[\/{4\a^2_2
A_o}{\a_1\sqrt{(t+\a_1)}}+\/{4\a^2_2
A_o}{\sqrt{\a_1}}\/{\arctgh\[\sqrt{1+\/{t}{\a_1}}\]}{
(t+\a_1)}\]\,,\cr
W(t)=&-\displaystyle\/{2}{\sqrt{\a_1}}\arctgh\[\sqrt{1+\/{t}{\a_1}}\].\nonumber
\end{align}\noindent For $\g=2$:\begin{align}
R(t)&=\displaystyle\left(1+\/{\a_1}{t}\right)^{8\a_2^2A_o/\a_1^3}
\exp\[-\/{4\a^2_2
A_o}{\a_1^2}\left(\/{2+\ln\[1+\/{\a_1}{t}\]}{(t+\a_1)}\right)\],\cr
W(t)&=\displaystyle\/{1}{\a_1^2}\ln\[1+\/{\a_1}{t}\]-\/{1}{\a_1
t}\,.\nonumber
\end{align}
And otherwise for $\g\neq{3}/{2},\,2$:\begin{align}
W(t)&=\/{\a_1^{(1-\g)}t^{(3-2\g)}}{(3-2\g)}{_2F_1}\!\left(\!3\!-\!2\g,\g,
4\!-\!2\g,\/{-t}{\a_1}\!\right)\!+\!
\displaystyle\/{\a_1^{-\g}t^{(4-2\g)}}{(4-2\g)}{_2F_1}\!\left(\!4-2\g,\g,
5\!-\!2\g,\/{-t}{\a_1}\!\right)\!,\cr
R(t)&=\int^t\!\!\/{W(s)}{(s+\a_1)^{2}}{\df s}\,,\nonumber
\end{align}
where $_2F_1$ denotes the hypergeometric function of the second
kind. The expression for the function $\theta$ is given by \beq
\theta(t)=\theta_o-\/{R(t)}{2\a_2}+\/{[5\a_1+W_o-2\a_2 z]}{2\a_2
(t+\a_1)}+\/{\a_2{X^2_o}}{(t+\a_1)}\big(1+\ln[t+\a_1]\big)\,,
\qquad\theta_o\in\R\,.\eeq Solution $G_2$ is always real and
non-singular if $t>0$, $\a_1>0$ and $\a_2\neq0$, and tends
asymptotically to zero when $t\rightarrow\infty$. It describes a
nonstationary and compressible flow in $(3+1)$ dimensions with
vorticity that lies in the $x$-$y$ plane. The current density
induced by \Ref{3.9} is equal to\beq \vJ=-\/{[\a_2\H_2+
\H_1]}{(t+\a_1)}\,\ve_1+\/{[\a_2\H_1- \H_2]}{(t+\a_1)}\ve_2\,,\eeq
and the resulting Lorentz force takes the form \beq \Fm=-\/{\a_2
X^2_o}{(t+\a_1)^2}\,{\rho}\,\ve_3\,.\eeq Since $\Fm$ is a
conservative force (\ie\,it can be derived from the gradient of the
magnetic pressure $|\vH|^2/2$) implies, by virtue of Kelvin's
theorem, that the circulation of the fluid is conserved.
\subsection*{$\boldsymbol{G_3=\big\{J_3+P_3+\a_1 G+\a_2
H,\,K_1,\,K_2\big\}}$,\quad $\a_1\neq 0$, $\a_2\in\R$} From this
algebra, we have computed the corresponding analytical solution
\begin{align}
\rho&=\/{R_o}{W}t^{2(\a_1-\a_2)/\a_1}\exp\[2\left(\/{\a_2}{\a_1}-1\right)
{\cal F}(s)\],\quad
p=\/{A_o}{W^\g}t^{-2\a_2/\a_1}\exp\[2\left(\/{\a_2}{\a_1}-\g\right){\cal
F}(s)\],\cr \v_1&=\/{x}{t}-\/{U_o}{t}\sin\big[\theta(s)\big],
\qquad\v_2=\/{y}{t}-\/{U_o}{t}\cos\big[\theta(s)\big],\qquad
\v_3=\/{1}{t}\big[W-{1}/{\a_1}\big]\,,\cr
{\H}_1&=\/{X_o}{W}t^{-\a_2/\a_1}\exp
\[\left(\/{\a_2}{\a_1}-\/{1}{\a_1}\right){\cal F}(s)\]
\sin\big[\theta(s)\big]\,,\cr {\H}_2&=\/{X_o}{W}t^{-\a_2/\a_1}
\exp\[\left(\/{\a_2}{\a_1}-\/{1}{\a_1}\right){\cal F}(s)\]
\cos\big[\theta(s)\big],\qquad{\H}_3=0\,, \label{3-91a}
\end{align}
where $R_o,\,A_o,\,U_o,\,V_o$ and $X_o$ are arbitrary constants. The
functions ${\cal F}$ and $\theta$ are given by \beq{\cal
F}=\int^s\/{\df s'}{W}\,,\quad\qquad\theta=\theta_o+\/{\ln
t}{\a_1}-\/{{\cal F}}{\a_1}\,,\qquad\qquad\theta_o\in\R\,,\eeq which
both depend on the symmetry variable $s=z+(\ln t)/{\a_1}$. The
unknown function $W$ is determined by solving the reduced form of
the equation \Ref{s2} along the $z$-axis:
\begin{align}
W^3\ds{W}-W^3+\/{W^2}{\a_1}&-\/{A_o}{R_o}\[\g\ds{W}+2
\left(\g-\/{\a_2}{\a_1}\right)\]W^{(2-\g)}\,\exp\big[2(2-\g){\cal
F}(s)\big]\cr
&-\/{X^2_o}{R_o}\[\ds{W}+\left(\/{1}{\a_1}-\/{\a_2}{\a_1}\right)\]\exp\[\left(4-\/{2}{\a_1}
\right){\cal F}(s)\]=0\,.\label{int91-i}
\end{align}
Equation \Ref{int91-i} is difficult to solve completely. However,
with the ansatz, \beq W=C_1 s+C_2,\;\quad\; C_1,\;C_2\in\R,\;\quad\;
C_1\neq 0\,,\label{ansatz}\eeq we can transform \Ref{int91-i} into a
polynomial equation \beq (C_1-1)W^3+\/{W^2}{\a_1}-\/{A_o}{R_o}\[(
C_1+2)\g-\/{2\a_2}{\a_1}\]W^{m}-\/{X^2_o}{R_o}
\[C_1+\left(\/{1}{\a_1}-\/{\a_2}{\a_1}\right)\]
W^{n}=0\,.\nonumber\eeq where $m=(2-\g)(C_1+2)/C_1$ and
$n=2(2\a_1-1)/(\a_1 C_1)$. By equating the coefficients of the same
power of $W$, we obtain five cases of invariant solutions of Eqs
\Ref{s1}--\Ref{s5}.
\vskip0.2in
\par
{\bf Case 1.)}\quad $C_1=1$,\qquad $\g=4/3$,\qquad$\a_2=\a_1+1$
\begin{align}
\rho&=R_o t^{-2/\a_1}\,W_1^{(2-3\a_1)/\a_1},\quad p=\/{R_o
}{2(\a_1-1)}t^{-2(\a_1+1)/\a_1}\,W_1^{2(1-\a_1)/\a_1},\label{3-911a}\\
\v_1&=\/{x}{t}-\/{U_o}{t}\sin\theta_1,\qquad
\v_2=\/{y}{t}-\/{U_o}{t}\cos\theta_1,\qquad\v_3=\/{1}{t}
\[W_1-\/{1}{\a_1}\]\,,\cr {\H}_1&=X_o\,
t^{-(\a_1+1)/\a_1}\sin\theta_1,\qquad {\H}_2=X_o\,
t^{-(\a_1+1)/\a_1}\cos\theta_1,\qquad {\H}_3=0\,,\nonumber
\end{align}
where $R_o\in\R^{+}$, $U_o$, $X_o\in\R$; $\a_1>0$. The functions
$W_1$ and $\theta_1$ are given by \beq W_1=\/{\ln
[t]}{\a_1}+z+C_2,\qquad\quad \theta_1=\theta_o-\/{1}{\a_1}\ln
\[\/{W_1}{t}\]\,,\qquad C_2,\,\theta_o\in\R\,.\nonumber\eeq
\par
{\bf Case 2.)}\quad $C_1=1$,\qquad $\g=2\a_2/3$,\qquad $\a_1=1$
\begin{align}
\rho&=R_o\,t^{2(1-\a_2)}\,W_2^{(2\a_2-5)},\qquad
p={A_o}\,t^{-2\a_2}\,,\label{3-911b}\\
\v_1&=\/{x}{t}-\/{U_o}{t}\sin\theta_2,\qquad
\v_2=\/{y}{t}-\/{U_o}{t}\cos\theta_2,\qquad\v_3=\/{1}{t}
\big[W_2-1\big]\,,\cr {\H}_1&=\/{\sqrt{R_o}}{\sqrt{(2-\a_2)}}\,
t^{-\a_2}\,W_2^{(\a_2-2)}\sin\theta_2\,,\cr {\H}_2&=\/{\sqrt{
R_o}}{\sqrt{(2-\a_2)}}\,t^{-\a_2}W_2^{(\a_2-2)}\cos\theta_2,\qquad
{\H}_3=0\,,\nonumber
\end{align}
where $R_o$, $A_o\in\R^{+}$, $U_o$, $X_o\in\R$; $3/2\leq\a_2<2$. The
functions $W_2$ and $\theta_2$ take the form \beq W_2=\ln
[t]+z+C_2,\qquad\theta_2=\theta_o-\ln
\[\/{W_2}{t}\]\qquad C_2,\,\theta_o\in\R\,.\nonumber\eeq
\par
{\bf Case 3.)}\quad $C_1=1$,\qquad $\g=4/3$,\qquad $\a_1=1$
\begin{align}
\rho&=(2-\a_2)\[2A_o+X^2_o\]t^{2(1-\a_2)} \,W_2^{(2\a_2-5)},\qquad
p=A_o
t^{-2\a_2}\,W_2^{(2\a_2-4)}\,,\label{3-911c}\\
\v_1&=\/{x}{t}-\/{U_o}{t}\sin\theta_2,\qquad
\v_2=\/{y}{t}-\/{U_o}{t}\cos\theta_2,\qquad
\v_3=\/{1}{t}\big[W_2-1\big]\,,\cr
{\H}_1&={X_o}{t^{-\a_2}}\,W_2^{(\a_2-2)}\sin\theta_2,\qquad
{\H}_2={X_o}{t^{-\a_2}} \,W_2^{(\a_2-2)}\cos\theta_2,\qquad
{\H}_3=0\,,\nonumber
\end{align}
where $A_o\in\R^{+}$, $U_o$, $X_o\in\R$ ; $1<\a_2<2$. The functions
$W_2$ and $\theta_2$ retain the form given in solution \Ref{3-911b}.
\par
{\bf Case 4.)}\quad $C_1=\displaystyle{(2\a_1-1)/{\a_1}}$\,,\qquad
$\g=(2\a_1+1)/(4\a_1-1)$
\begin{align}
\rho&=R_o\,
t^{2(\a_1-\a_2)/\a_1}\,W_3^{(2\a_2-6\a_1+1)/(2\a_1-1)}\,,
\label{3-911-d}\\
p&=\/{(1-\a_1)R_o}{[2(\a_2-\a_1)-1]}\,t^{-2\a_2/\a_1}
W_3^{(2\a_2-2\a_1-1)/(2\a_1-1)}\,,\cr
\v_1&=\/{x}{t}-\/{U_o}{t}\sin\theta_3,\qquad
\v_2=\/{y}{t}-\/{U_o}{t} \cos\theta_3,\qquad
\v_3=\/{1}{t}\[W_3-\/{1}{\a_1}\]\,,\cr {\H}_1&=\/{\sqrt
{R_o}}{\sqrt{(2\a_1-\a_2)}}\,{t^{-\a_2/\a_1}}W_3^{(\a_2-2\a_1)/(2\a_1-1)}\sin
\theta_3\,,\cr {\H}_2&=\/{\sqrt
{R_o}}{\sqrt{(2\a_1-\a_2)}}\,{t^{-\a_2/\a_1}}W_3^{(\a_2-2\a_1)/(2\a_1-1)}
\cos\theta_3,\qquad {\H}_3=0\,,\nonumber
\end{align}
where $R_o\in\R^{+}$, $U_o\in\R$. The parameters $\a_1$ and $\a_2$
lie in these intervals: $1/2<\a_1<1$, $2\a_1<\a_2<(2\a_1+1)/2$. The
functions $W_3$ and $\theta_3$ take the form \beq
W_3=\/{(2\a_1-1)}{\a_1}\[\/{\ln[t]}{\a_1}+z\]+C_2,\qquad
\theta_3=\theta_o+\/{\ln[t]}{\a_1}+\/{\ln [W_3]}{(1-2\a_1)},\quad
C_2,\,\theta_o\in\R\,.\nonumber\eeq
\par
{\bf Case 5.)}\quad $C_1=\displaystyle{(4\a_1-2)}/{3\a_1}$\,,\qquad
$\g=6\a_1/(5\a_1-1)$
\begin{align}
\rho&=R_o\,
t^{2(\a_1-\a_2)/\a_1}W_4^{(3\a_2-8\a_1+1)/(2\a_1-1)}\,,\label{3-911-e}\\
p&=\/{R_o }{2(2\a_1-\a_2)}t^{-2\a_2/\a_1}
W_4^{3(\a_2-2\a_1)/(2\a_1-1)}\,,\cr
\v_1&=\/{x}{t}-\/{U_o}{t}\sin\theta_4,\quad \v_2=\/{y}{t}-\/{U_o}{t}
\cos\theta_4,\quad \v_3=\/{1}{t}\[W_4-\/{1}{\a_1}\]\,,\cr
{\H}_1&=\sqrt{\/{ (\a_1-2)R_o}{(4\a_1-3\a_2+1)}}\,{t^{-\a_2/\a_1}}
W_4^{(3\a_2-4\a_1-1)/(4\a_1-2)}\sin\theta_4\,,\cr {\H}_2&=\sqrt{\/{
(\a_1-2)R_o}{(4\a_1-3\a_2+1)}}\,{t^{-\a_2/\a_1}}W_4^{(3\a_2-4\a_1-1)/(4\a_1-2)}
\cos\theta_4,\quad {\H}_3=0\,,\nonumber
\end{align}
where $R_o\in\R^{+}$; $U_o\in\R$. Here $\a_1$ and $\a_2$ lie in:
$1/2<\a_1<2$ and $(4\a_1+1)/3<\a_2<2\a_1$, $\a_1>2$ and
$\a_1<\a_2<(4\a_1+1)/3$. The functions $W_4$ and $\theta_4$ are
given by\beq W_4=\/{(4\a_1-2)}{3\a_1}\[\/{\ln[t]}{\a_1}
+z\]+C_2,\quad
\theta_4=\theta_o+\/{\ln[t]}{\a_1}+\/{3}{(2-4\a_1)}\ln [{W_4}],\quad
C_2,\,\theta_o\in\R\,.\nonumber\eeq
Solutions \Ref{3-911a}--\Ref{3-911-e} are real and non-singular if
$W_i>0$ ($i=1,2,3,4$) and $t>0$, they all tend asymptotically to
zero for $t\rightarrow\infty$. These solutions describe a
nonstationary and compressible flow with vorticity that lies in the
$x$-$y$ plane. Note that for the solution \Ref{3-911b} there is no
gradient of fluid pressure. Except for the solution \Ref{3-911a}
where $\Fm=\boldsymbol{0}$, applying the expression for the Lorentz
force \Ref{em} we find than that only the first term (\ie\,the
magnetic pressure) yields a nonzero contribution, while the second
term (the magnetic tension force) vanishes. So, $\Fm$ is a
conservative force that acts only along the $z$-axis as explained
for solutions $G_1$ and $G_2$. Consequently, by virtue of Kelvin's
theorem, the circulation of the fluid is constant for solutions
\Ref{3-911a}--\Ref{3-911-e}.
\section{$\boldsymbol{\vH=\big({\H}_1,{\H}_2, {\H}_3\big)}$}\label{sec2}
\subsection*{$\boldsymbol{G_4=\big\{F+\a_1 G+\a_2 H,\,P_o,\,P_3\big\}}$,
\qquad $\a_1\neq 0,1,\quad\a_2\in\R$} By choosing  $\a_2=-1$, we
were able to solve completely the reduced system and we get
\begin{align}
\rho&=x^{-2(1+\a_1)}R\,(s),\;\qquad p=A_o-\/{(c_o^2
Y_o^2+Z_o^2)}{2(x^2+y^2)}\csc^{2}\big(c_o\theta\big)\,,\cr
\v_1&=0,\qquad \v_2=0,\qquad\,
\v_3=W_o\,(x^2+y^2)^{\a_1/2}\,\sin^{\a_1}\big(c_o\theta\big)\,,\cr
\H_1&=\/{c_o Y_o\,
x}{(x^2+y^2)}\cot(c_o\theta)+\/{Y_o\,y}{(x^2+y^2)}\,,\label{g4}\\
\H_2&=\/{c_o
Y_o\,y}{(x^2+y^2)}\cot\big(c_o\theta\big)-\/{Y_o\,x}{(x^2+y^2)},\qquad
\H_3=\/{Z_o}{\sqrt{x^2+y^2}}\csc\big(c_o\theta\big)\,,\nonumber
\end{align}
where $A_o\in\R^{+}$; $c_o$, $W_o$, $Y_o$, $Z_o\in\R$, with $c_o\neq
0$. As a result of the incompressibility of the fluid, $R$ is an
arbitrary function (but positively defined) of the symmetry variable
$s=x/y$. The function $\theta$ is given by
\beq\theta=\arctan\left(\/{y}{x}\right).\eeq Notice that for a given
$c_o\in\N/\{0\}$, the trigonometric functions can be expanded in
terms of rational functions of $x$ and $y$.  This solution is
non-singular if $x,y\neq 0$, and $y\neq x\tan\big[k\pi/c_o\big]$,
$k\in\Z$. We note that this solution does not depend on the
variables $z$ and $t$: the axis of symmetry coincides with the
$z$-axis which is also the direction of the stationary flow (the
paths of fluid elements are streamlines). Solution \Ref{g4}
describes a static equilibrium [\cf\,the steady-state form of
\Ref{s2}]: \beq\grad p=\vJ\times\vH\,,\label{ee1}\eeq where the
current density $\vJ=(\J_1,\J_2,\J_3)$, given by the Amp\`ere's law,
takes the form \beq \J_1=-\/{\H_1}{Y_o}\H_3,\qquad
\J_2=-\/{\H_2}{Y_o}\H_3,\qquad
\J_3=\/{c_o^2\,Y_o}{(x^2+y^2)}\csc^2\big(c_o\theta\big)\,.\eeq
Nothing that $\vJ\times\vH=-\grad(|\vH|^2/2)+(\vH\cdot\nabla)\vH$,
so to achieve equilibrium the magnetic tension must balance not only
the magnetic pressure gradient but that of the fluid pressure as
well. Also, we have $(\vH\cdot\nabla)\vv=\boldsymbol{0}$ which
indicates that the flow velocity $\vv$ remains constant along $\vH$.
The flow has vorticity given by $\vw=(\w_1,\w_2,0)$, with \beq
\w_1=\a_1\/{\big[x+c_o
y\cot(c_o\theta)\big]}{(x^2+y^2)}\,\v_3,\qquad\w_2=\a_1\/{\big[y-c_o
x\cot(c_o\theta)\big]}{(x^2+y^2)}\,\v_3\,.\eeq By virtue of Kelvin's
theorem the circulation of the fluid is preserved because
$\df\vv/\df t=\boldsymbol{0}$ [\cf\,equation \Ref{vort} is
identically satisfied].

The solution \Ref{g4} can be interpreted physically as the
propagation of a stationary double entropic wave $\rm{E}_1\rm{E}_1$
(resulting from the nonlinear superposition of two entropic simple
waves $\rm{E}_1$) in an incompressible fluid \cite{pic}. In the next
two solutions, we present such type of wave in the context of
cylindrically and spirally magnetic geometry.
\subsection*{$\boldsymbol{G_5=\big\{J_3+\a_1 G+\a_2 H,\,P_o,\,P_3\big\}}$,
\qquad $\a_1$,\;\;$\a_2\in\R$} From this algebra, we get the
following invariant solution in cylindrical coordinates
$(r,\varphi,z)$
\begin{align}
\rho&=\exp[2(\a_2-\a_1)\varphi]R(r),\qquad\;
p=A_o\,\exp\big[2\a_2\varphi-2\a_2\vartheta(r)\big]\,,\label{g5}\\
\v_r&=0,\qquad \v_\varphi=0,\qquad
\v_z=W_o\,\exp\big[\a_1\varphi-\a_2\vartheta(r)\big]\,,\cr
\H_r&=\/{X_o}{r} \exp\big[\a_2\varphi-\a_2\vartheta(r)\big]\,,\cr
{\H}_\varphi&=\/{X_o}{r}
\exp\big[\a_2\varphi-\a_2\vartheta(r)\big]\tan\big[Y(r)\big],\quad
\H_{z}=Z_o\,\exp\big[\a_2\varphi-\a_2\vartheta(r)\big]\,,\nonumber
\end{align}
where $A_o\in\R^{+}$, $W_o$, $X_o$, $Z_o\in\R$; $R$ is an arbitrary
function (with $R>0$) of the symmetry variable $s=r$ (its level
surfaces represent cylinders), and the function $\vartheta$ is given
by \beq\vartheta=\int^r\!\!\tan (Y)\/{\df r'}{r'}\,.\eeq The
function $Y$ is a solution of the nonlinear first order ODE \beq
r\/{\df Y}{\df r}-\a_2\b_o r^2\,\cos^2(Y)-\a_2=0\,,\label{cyl}\eeq
where the parameter $\b_o$ is defined by
\beq\b_o=\/{(2A_o+Z^2_o)}{X^2_o}\,.\label{param}\eeq Equation
\Ref{cyl} does not have the Painlev\'e property, and is difficult to
integrate. If such an equation has the  Painlev\'e property, \ie\,
if its general solution has no ``movable'' singularities other than
poles (such as essential singularities or branch points), then it
can be transformed into one of transcendents forms (Ince
\cite{ince}) and integrated in terms of some known functions. The
test verifying whether a given ODE satisfies certain necessary
condition for having the Painlev\'e property, is algorithmic
\cite{ablo} and can be performed using a specifically written
MATHEMATICA program \cite{bald}. A complete analysis of this ODE
would take us beyond the scope of this paper. However, we can
discuss some interesting physical properties of this solution.

The Amp\`ere's law shows that three components of current, $\J_r$,
$\J_\varphi$ and $\J_z$, are nonzero: \beq
\J_r=\/{\a_2}{r}\H_z,\quad \J_\varphi=\/{\a_2}{r}\tan(Y)\H_z,\quad
\J_z=\/{\H_r}{\cos^2(Y)}\/{\df Y}{\df r} -\/{\a_2}{r}
\H_\varphi\tan(Y)-\/{\a_2}{r}\H_r\,.\eeq For $\a_2=0$, the current
density $\vJ$ is identically zero. So $\rot\vH=\boldsymbol{0}$
permits the expression of the magnetic field as the gradient of a
magnetic scalar potential, $\vH=-\grad\phi_M$.

Notice that the equation \Ref{cyl} corresponds to the reduced form
of the steady-state of equation \Ref{s2} along both the $r$-and
$\varphi$-axes:\beq\Dp{}{r}\big[p+|\vH|^2/2\big]=(\vH\cdot\nabla)_r\,\H_r,
\qquad\;\;\/{1}{r}\Dp{}{\varphi}\big[p+|\vH|^2/2\big]=
(\vH\cdot\nabla)_\varphi\,\H_\varphi\,,\label{equ-cyl}\eeq which
both describe a balance between the pressure gradient $\grad p$ and
the Lorentz force $\Fm$. The two terms on the left side of the
equalities represent the fluid pressure and the magnetic pressure,
while the terms on the right side represent the tension forces
generated by the curvature of the magnetic field lines. Both the
magnetic field $\vH$ and current density $\vJ$ lie on constant
pressure surfaces since $\vH\cdot\nabla p=0$ and $\vJ\cdot\nabla
p=0$.

Solution $G_5$ is non-singular when $Y\neq (2k+1)\pi/2$, $k\in\Z$,
and $r\neq 0$. This solution is independent of $z$ and $t$. The
solution $G_5$ represents an equilibrium two-dimensional
configuration for which the stationary and incompressible flow is
along the symmetric axis (\ie\,the $z$-axis) of a cylindrically
symmetric magnetic field $\vH$. Also,
$(\vH\cdot\nabla)\vv=\boldsymbol{0}$ confirms that $\vv$ remains
constant along $\vH$. The flow has a vorticity field given by
\beq\vw=\left(\/{\a_1}{r}\v_z,\,
\/{\a_1}{r}{\tan(Y)}\v_z,\,0\right)\,,\eeq and the circulation of
the fluid is preserved since $\df\vv/\df t=\boldsymbol{0}$.

Finally, if we impose the condition that
\beq\vH\cdot\nabla\rho=0\,,\label{const-cyl}\eeq  then we can
determine the explicit form of the
density:\beq\rho=R_o\exp\big[2(\a_2-\a_1)\varphi-2\a_2\vartheta(r)\big],\qquad\qquad
R_o\in\R^{+}\,.\eeq Equation \Ref{const-cyl} simply states that the
lines of force lie on the surfaces $\rho=\rm{const}$ which coincide
with the isobaric surfaces. So, any given smooth equilibrium
function $\vH$, $\vv$, $\rho$ and $p$ in $\R^3$ defines a
distribution of magnetic surfaces $\psi=\rm{const}$ in $\R^3$
\cite{moffat}.
\subsection*{$\boldsymbol{G_6=\big\{J_3+\a_1(F+G)+\a_2 H,\,P_o,\,P_3\big\}}$,\qquad
$\alpha_1\neq 0,\;\;\a_2\in\R$} In cylindrical coordinates
$(r,\varphi,z)$, the corresponding GIS to this algebra has the form
\begin{align}
\rho&=\exp\big[2(\a_2-\a_1)\varphi\big]\,R(s),\quad\qquad
p=A_o\,\exp\big[2\a_2\varphi-2\a_2\vartheta_2(s)\big]\,,\label{g6}\\
\v_r&=0,\quad\qquad \v_\varphi=0,\qquad\;\,
\v_z=W_o\,\exp\big[\a_1\varphi-\a_2\vartheta_2(s)\big]\,,\cr
{\H}_r&=X_o\/{\exp\big[\a_2\varphi
-\vartheta_1(s)-\a_2\vartheta_2(s)\big]}
{\cos[Y(s)]-\a_1\sin[Y(s)]}\cos[Y(s)] \,,\cr
{\H}_\varphi&=X_o\/{\exp\big[\a_2\varphi
-\vartheta_1(s)-\a_2\vartheta_2(s)\big]}
{\cos[Y(s)]-\a_1\sin[Y(s)]}\sin[Y(s)],\qquad
{\H}_z=Z_o\,\big[\a_2\varphi-\a_2\vartheta_2(s)\big]\,,\nonumber
\end{align}
where $A_o\in\R^{+}$,  $W_o$, $X_o$, $Z_o\in\R$; and $R$ is an
arbitrary function (but positively defined) of the symmetry variable
$s=re^{-\a_1\varphi}$. The level surfaces of the symmetry variables
are logarithmic spirals. The functions $\vartheta_1$ and
$\vartheta_2$ are given by \beq\vartheta_1=\int^s\!\!\/{\cos
(Y)}{\big[\cos(Y)-\a_1\sin(Y)\big]}\/{\df s'}{s'},\qquad
\vartheta_2=\int^s\!\!\/{\sin
(Y)}{\big[\cos(Y)-\a_1\sin(Y)\big]}\/{\df s'}{s'}\,.\eeq The
functions $Y$ solves the integro-differential equation: \beq s\/{\df
Y}{\df s}-\/{\a_2\b_o }{(1+\a_1^2)}\,[\cos(Y)-\a_1\sin(Y)]^2
\exp\[2\vartheta_1\,(s)
\]-\/{(\a_1+\a_2)}{(1+\a_1^2)}=0\,,\label{heli}\eeq
where $\b_o$ retains the form \Ref{param}, and with the conditions
that $Y\neq\arctan(1/\a_1)+k\pi$, $k\in\Z$, and $s\neq 0$, which
ensure that the solution $G_6$ is non-singular. This solution does
not depend on the variables $z$ and $t$. Equation \Ref{heli}
corresponds to the same reduced form of $\grad p=\vJ\times\vH$ along
both the $r$- and $\varphi$-axes [\cf\,equation \Ref{equ-cyl}]. The
determination of the general solution of equation \Ref{heli} is a
difficult task. We have not able to find an explicit solution of
\Ref{heli} and we must look for a solution by numerical methods.
Nevertheless, we can resume some physical properties of this
solution.

The solution $G_6$ represents a two-dimensional spirally symmetric
configuration in static equilibrium; the steady flow is parallel the
axis of symmetry (\ie\,the $z$-axis), and is conserved along the
magnetic field lines since $(\vH\cdot\nabla)\vv=\boldsymbol{0}$. The
vorticity of the flow is given by $\vw=(\w_r,\w_\varphi,0)$, with
\beq\w_r=\/{\a_1}{r}\/{\cos(Y)\v_z}{\big[\cos(Y)-\a_1\sin(Y)
\big]}\,,\qquad
\w_\varphi=\/{\a_1}{r}\/{\sin(Y)\v_z}{\big[\cos(Y)-\a_1\sin(Y)
\big]}\,,\eeq and the circulation of the fluid is preserved since we
have $\df\vv/\df t=\boldsymbol{0}$.

The current induced by $\vH$ has three nonzero components $\J_r$,
$\J_\varphi$ and $\J_z$, given by:
\begin{align}
\J_r&=\/{\a_2}{r}\/{\cos(Y)\H_z}{[\cos(Y)-\a_1\sin(Y)]},\quad\qquad
\J_\varphi=\/{\a_2}{r}\/{\sin(Y)\H_z}{[\cos(Y)-\a_1\sin(Y)]}\,,\cr
\J_z&=\/{1}{r}\big(\H_\varphi-\a_2 \H_r\big)
-\/{\big[\cos(Y)+\a_2\sin(Y)\big]}{r\big[\cos(Y)-\a_1\sin(Y) \big]}
\big(\H_\varphi+\a_1 \H_r\big)\cr
&+(1+\a_1^2)\,\exp[-\a_1\varphi]\,\/{\big[\cos(Y)\H_r+\sin(Y)\H_\varphi\big]}
{\big[\cos(Y)-\a_1\sin(Y)\big]}\/{\df Y}{\df s}\,.
\end{align}
If we impose the condition that $\vH\cdot\nabla\rho=0$, then we get
the expression for $\rho$
\beq\rho=R_o\exp\big[2(\a_2-\a_1)\varphi-2\a_2\vartheta_2(s)\big],\qquad\qquad
R_o\in\R^{+}\,.\eeq So $\vH$, $\vv$, $\rho$ and $p$ define now a
distribution of magnetic surfaces $\psi=\rm{const}$ in $\R^3$.

It is noteworthy that for $\a_2=0$, the equation \Ref{heli} can be
easily solved to give:
\begin{align}
\rho&=\exp\big[-2\a_1\varphi\big]R(s),\qquad p=A_o\,;\qquad
\v_{r}=0,\qquad\v_{\varphi}=0\,,\cr \quad
v_z&=W_o\,r^{\a_1/(\a_1^2+1)}\,\exp\[\/{\a_1}{(\a_1^2+1)}\,\varphi\]
\big[\a_1\sin(Y)-\cos(Y)\big]\,,\\ \H_{r}&={X_o}\,
r^{-1/(\a_1^2+1)}\,\exp\[\/{\a_1}{(\a_1^2+1)}\,\varphi\]\cos(Y)\,,\cr
\H_{\varphi}&={X_o}\,
r^{-1/(\a_1^2+1)}\,\exp\[\/{\a_1}{(\a_1^2+1)}\,\varphi\]\sin(Y)\,,\qquad
\H_z=Z_o\,,\nonumber
\end{align}
where $A_o\in\R^{+}$, $\a_1\neq 0$, $W_o$, $X_o$, $Z_o\in\R$; and
$R$ is an arbitrary function (with $R>0$) of $s=re^{-\a_1\varphi}$.
The expression for the function $Y$ is given by \beq
Y=Y_o+\/{\a_1}{(\a_1^2+1)}\ln
[r]-\/{\a^2_1}{(\a_1^2+1)}\,\varphi\,.\eeq Here we have an example
of a two-dimensional configuration with spirally symmetric magnetic
geometry. Since $\vJ=\boldsymbol{0}$, the magnetic field $\vH$ is
potential.
\subsection*{$\boldsymbol{G_7=\big\{J_3+P_3,\,P_1,\,P_2\big\}}$}
The reductions of this algebra lead us to this invariant solution of
physical interest:
\begin{align}
\rho&=\rho_o,\quad\qquad p=p_o\,,\label{3.14}\\
\v_1&=\/{U(t)}{\sqrt{\rho_o}}\sin\big[V(t)-z\big],\qquad
\v_2=\/{U(t)}{\sqrt{\rho_o}}\cos\big[V(t)-z\big],\qquad \v_3=W_o\,,\nonumber\\
{\H}_1&=X(t)\sin\big[Y(t)-z\big],\qquad
{\H}_2=X(t)\cos\big[Y(t)-z\big],\qquad {\H}_3=Z_o\,,\nonumber
\end{align}
where $\rho_o$, $p_o\in\R^{+}$; $W_o,\,Z_o\in\R$ and $Z_o\neq 0$.
The functions $U$, $V$, $X$ and $Y$ have the form
\begin{align}
U(t)&=\[E_o-\sqrt{E^2_o-\rho_o\delta_o^2}\sin\left(\phi_o-\/{2
Z_o}{\sqrt{\rho_o}}t\right)\]^{1/2},\,\cr
X(t)&=\[E_o+\sqrt{E^2_o-\rho_o\delta_o^2}\sin\left(\phi_o-\/{2
Z_o}{\sqrt{\rho_o}}t\right)\]^{1/2}\,,\cr V(t)&=V_o+W_o\,
t+\arctan\[\/{E_o}{\sqrt{\rho_o}\,\delta_o}\tan\left(\/{\phi_o}{2}
-\/{Z_o}{\sqrt{\rho_o}}t\right)-\/{\sqrt{E^2_o-\rho_o\delta_o^2}}
{\sqrt{\rho_o}\,\delta_o}\]\,,\cr
Y(t)&=V+\arccos\[\/{\sqrt{\rho_o}\,\delta_o}{UX}\]\,,\nonumber
\end{align}
where $\phi_o,\,V_o$ and $Y_o\in\R$. The constant $E_o$ is equal to
 \beq E_o=\/{\rho_o}{2}[U(t)]^2+\/{1}{2}[X(t)]^2\,,\label{csv-14} \eeq which
 corresponds to the sum of kinetic and magnetic energies
 perpendicular to the $z$-axis. The constant $\delta_o$ is given by
\beq \delta_o=\v_1\H_1+\v_2\H_2\,,\label{coll-14}\eeq with
$\d^2_o\leq E^2_o/\rho_o$ ensuring that the solution $G_7$ is always
real. Thus, the longitudinal component of $\vv$ along the magnetic
field $\vH$ is a constant quantity.

The solution \Ref{3.14} describes, along with the relations
\Ref{csv-14} and \Ref{coll-14}, the main properties of the double
Alfv\'en-entropic wave $\rm{A}\rm{E}_1$ which results from the
nonlinear superposition of an Alfv\'en wave $\rm{A}$ with an
entropic wave $\rm{E}_1$ \cite{pic}. Such waves of an arbitrary
amplitude were found for the first time by Alfv\'en \cite{alf1} as
nonstationary solutions of MHD equations for an incompressible
medium (producing no density or pressure fluctuations). The wave
motion can be explained by the intrinsic nature of the Lorentz force
\beq
\Fm=(\vH\cdot\nabla)\vH=-Z_o\H_2\,\ve_1+Z_o\H_1\,\ve_2\label{fm-14}\,,\eeq
which corresponds to tension force along the field lines of $\vH$.
So that the magnetic field lines twist relatively to one another,
but do not compress. Namely, the double wave $\rm{A}\rm{E}_1$
modifies the flow direction that is not parallel to the magnetic
field $\vH$. The flow has vorticity that lies in the $x$-$y$ plane,
and according to the Kelvin's theorem, the circulation of the fluid
is not conserved because of the presence of tension forces
originating from $\Fm$. Note that the instantaneous power dissipated
by the magnetic force is given by
\beq\Fm\cdot\vv=\/{\sqrt{E^2_o-\rho_o\delta_o^2}}
{\sqrt{\rho_o}}\cos\left(\phi_o-\/{2
Z_o}{\sqrt{\rho_o}}t\right)\,.\eeq Indeed, the equation \Ref{csv-14}
describes basic oscillation between perpendicular fluid kinetic
energy and perpendicular ``line bending'' magnetic energy, \ie\,a
balance between inertia and field lines tension. For the particular
case $E_o=\sqrt{\rho_o}|\delta_o|$; $\vv$ are $\vH$ collinear: there
is no coupling between flow and $\vH$. Finally, note that the
conservation law of energy \cite{landau}
\beq\Dp{}{t}\[\/{1}{2}\rho|\vv|^2+\/{p}{(\g-1)}+\/{|\vH|^2}{2}\]+
\div \left\{\[\/{1}{2}\rho|\vv|^2+\/{\g
p}{(\g-1)}+|\vH|^2\]\vv+(\vH\cdot\vv)\vH\right\}=0\,,\nonumber\eeq
turns out to be trivially satisfied.
\subsection*{$\boldsymbol{G_8=\big\{F+G,\,K_1+P_2+\a P_3,\,P_1+\b P_2\big\}}$,
\qquad$\a\in\R$,\quad$\b>0$} The corresponding invariant solution
has the explicit form
\begin{align}
\rho&=\/{R(t)}{[\a (\b x-y)+(1-\b t)z]^2},\qquad
p={A_o}{\big[R(t)\big]^{-\g}}\,,\label{g7}\\
\v_1&=\/{z}{\a}+\/{(W_o\,t-U_o)}{\a R(t)}{\big[\a (\b x-y)+(1-\b
t)z\big]}\,,\cr\v_2&=\/{V_o}{R(t)}\big[\a (\b x-y)+(1-\b
t)z\big],\quad \v_3=-\/{W_o}{R(t)}\big[\a (\b x-y)+(1-\b
t)z\big]\,,\cr {\H}_1&=\/{\[Z_o\,t+\a X_o\]}{\a R(t)},\qquad
{\H}_2=\/{Y_o}{R(t)},\qquad {\H}_3=\/{Z_o}{R(t)}\,,\nonumber
\end{align}
where $A_o$, $W_o\in\R^{+}$; $U_o$, $V_o$, $X_o$, $Y_o$, $Z_o\in\R$,
$\a\in\R/\{0\}$. The function $R(t)$ is given by \beq R(t)=\b
W_o\,t^2-[W_o+\b U_o+\a V_o] t+R_o\,,\eeq where the constant $R_o$
must satisfy \beq R_o>\/{\[W_o+\b U_o+\a V_o\]^2}{2\b W_o}\,,\eeq in
order to obtain $\rho>0$. This solution represents a nonstationary
and compressible flow in $(3+1)$ dimensions. Note that this solution
is singular when $t=[z+\a(\b x-y)]/\b z$ which coincides with a
stagnation point in $\R^4$ (\ie\,where $\rho\rightarrow\infty$,
$\vv=\boldsymbol{0}$), and tends asymptotically to zero for
sufficiently large $t$. The fluid is force-free since both the
gradient of fluid pressure and $\Fm$ are zero. Force-free conditions
are widely applicable in astrophysical environments because forces
other than electromagnetic are comparatively much smaller
\cite{somov}. The vorticity field of the flow has the form
$\vw=(\w_1,\w_2,\w_3)$, with
\begin{align}
\w_1&=\/{\[\b V_o\,t+\a
W_o-V_o\]}{R(t)},\quad\w_2=\/{\[R_o-U_o+\a^2\b W_o-\a V_o t\]}{\a
R(t)},\quad \w_3=\/{\[W_o t+\a\b V_o-U_o\]}{R(t)}\,,\nonumber
\end{align}
Consequently, by virtue of Kelvin's theorem, the circulation of the
fluid is preserved.
\subsection*{$\boldsymbol{G_9=\big\{F+K_3+\a H,\,P_2,\,P_3+\b P_1\big\}}$,\qquad
$\b>0$,\quad$\a\in\R$} The reductions of this algebra lead us to the
corresponding analytical solution:
\begin{align}
\rho&=\/{R_o}{f}t^{2\a}\exp\[-(1+2\a)\!\int^s\!\/{\df
s'}{f}\],\qquad
p=\/{A_o}{f^\g}t^{2\a}\exp\!\[-(2\a+\g)\!\int^s\!\/{\df s'}{f}\],\\
\v_1&=U(s),\qquad\quad\v_2=V_o,\qquad\quad\v_3=\ln[t]+W(s),\cr
 {\H}_1&=\b\/{Z_o}{f}t^{\a}\exp\!\[-(1+\a)\!\int^s\!\/{\df s'}{f}\] ,
 \;\;\quad
 {\H}_2=\/{Y_o}{f}t^{\a}\exp\!\[-(1+\a)\!\int^s\!\/{\df s'}{f}\],\cr
{\H}_3&=\/{Z_o}{f}t^{\a}\exp\!\[-(1+\a)\!\int^s\!\/{\df
s'}{f}\],\nonumber
\end{align}
where $R_o$, $A_o$, $V_o$, $Y_o$ and $Z_o$ are arbitrary constants.
Also, we have introduced the function \beq f=U(s)-\b
W(s)-s+\b\label{3-20a}\,,\eeq that depends on the symmetry variable
$s=\displaystyle\b\ln[t]+{(x-\b z)}/{t}$. The unknown functions $U$
and $W$ are obtained by solving the overdetermined system which
consists of \Ref{3-20a} and these two equations:
\begin{align}
&\b\ds{U}+\ds{W}+1=0\,,\label{3-20b}\\
f^3\ds{f}&+f^3-\b f^2-(1+\b^2)\/{A_o}{R_o}\[\g\ds{f}+\g+2\a\]
f^{2-\g}\exp\[(1-\g)\!\!\int^s\!\/{\df s'}{f}\]\cr
&-\/{(1+\b^2)}{R_o}\big[Y_o^2+(1+\b^2)Z_o^2\big]
\[\ds{f}+1+\a\]\exp\[-\int^s\!\/{\df s'}{f}\]=0\,.\label{3-20c}
\end{align}
We have not succeeded in solving completely this system. However,
with the ansatz \beq W=C_1 s+C_2,\;\quad\;
C_1,\;C_2\in\R,\;\quad\;C_1\neq 0\,,\eeq we can calculate $f$ from
equation \Ref{3-20c} by setting $C_1=-1/2$ and $\g=3$. Next, by
substituting this result into equations \Ref{3-20a} and \Ref{3-20b},
we determine $U$ and $W$. Finally, we obtain the following invariant
solution of Eqs \Ref{s1}--\Ref{s5}:
\begin{align}
\rho&=-\/{1}{2\b}(2\a+1)(1+\b^2)\big[Y_o^2+(1+\b^2)Z_o^2\big]
t^{2\a}\,\xi^{(4\a+1)}\,,\cr p&=-\/{1}{2\b}\/{(2\a+1)}{(4\a+3)}
\big[Y_o^2+(1+\b^2)Z_o^2\big]t^{2\a}
\,\xi^{(4\a+3)}\,,\label{3.20}\\
\displaystyle
\v_1&=\/{2\b}{(1+\b^2)}\ln\big[t^{{1}/{4}}\,\xi\big]+\/{(x-\b
z)}{2(1+\b^2)t}+\/{C_2+\b(C_3-1)}{(1+\b^2)},\;\quad\; \v_2=V_o\,,\cr
\v_3&=\/{2}{(1+\b^2)}\ln\big[t^{{(2+\b^2)}/4}\,\xi\big]+\/{\b(\b
z-x)}{2(1+\b^2)t}+\/{C_3+\b(\b-C_2)}{(1+\b^2)}\,,
\nonumber\\
{\H}_1&=\b Z_o\, t^{2\a}\,\xi^{(2\a+1)},\qquad{\H}_2=Y_o\, t^{2\a}\,
\xi^{(2\a+1)},\qquad {\H}_3=Z_o\,t^{2\a}\,\xi^{(2\a+1)}\,,\nonumber
\end{align}
where \beq \xi=C_2-\/{\b}{2}\ln[t]-\/{(x-\b z)}{2t}\,,\eeq and
$\b>0$, $C_2$, $C_3$, $V_o$, $Y_o$ and $Z_o\in\R$. Solution
\Ref{3.20} is always real and non singular if $\xi>0$ and $t>0$.
Note that this solution does not depend on the variable $y$. In
order that $\rho$ and $p>0$, we must have $-3/4<\a<-1/2$. This
solution describes a nonstationary and compressible flow in $(2+1)$
dimensions with vorticity along the axis of symmetry:
\beq\vw=\big[t\xi\big]^{-1}\,\ve_2\,,\eeq which is everywhere normal
to the plane of flow (the $x$-$z$ plane). The current density
induced by $\vH$ takes the form \beq \vJ=-(2\a+1)Z_o
\,t^{(\a-1)}\,\xi^{2\a}\,\Big(\,1,-\,(1+\b^2),1\,\Big)\,.\eeq Thus,
the Lorentz force related to solution \Ref{3.20} is given by \beq
\Fm=\/{\b}{(1+\b^2)}\/{\rho}{t}\Big(\,1,\,0,\,-\b\Big)\eeq and
produced no tension along the lines of force. Also, we have
$(\vH\cdot\grad)\vv=\boldsymbol{0}$ which implies that $\vH/\rho$ is
constant along the streamlines, and thus $\vH$ remains inextensible
[\cf\,\Ref{lines}]. So, along both the $x$- and $z$-axes the Lorentz
force $\Fm$ causes compressions and expansions of the distance
between the lines of force without changing their direction, exactly
as does the propagation of a double magnetoacoustic wave
$\rm{F}\rm{F}$ which results from the nonlinear superposition of two
magnetoacoustic fast waves $\rm{F}$ that propagate perpendicular to
each other in an ideal fluid \cite{pic}. Since $\Fm$ is a
conservative force (\ie\,it comes from the gradient of magnetic
pressure $|\vH|^2/2$), the circulation of the fluid \Ref{circ} is
conserved.
\subsection*{$\boldsymbol{G_{10}=\big\{G+\a H,K_1,K_2\big\}}$, \quad $
\a\in\R$} For $\a=2$, the reduced system corresponding to this
algebra admits this exact solution
\begin{align}
\rho&=\/{R_o}{t^2}W^{-1},\quad\quad p=\/{A_o}{t^4}
W^{-\g}\exp\[2(2-\g)\int^z\!
\/{\df z'}{W}\]\,,\nonumber\\
\v_1&=\/{x-U_o}{t}+\/{a^2_o}{Z_o}\/{X_o\,t^{-1}}{(W-a^2_o)}
\exp\[\int^z\!\!\/{\df z'}{(W-a^2_o)}\]\,,\\
\v_2&=\/{y-V_o}{t}+\/{a^2_o}{Z_o}\/{Y_o\,t^{-1}}{(W-a^2_o)}
\exp\[\int^z\!\!\/{\df z'}{(W-a^2_o)}\]\,,\quad\quad
\v_3=\/{W}{t}\,,\cr {\H}_1&=\/{
X_o\,t^{-2}}{(W-a^2_o)}\exp\[\int^z\!\!\/{\df
z'}{(W-a^2_o)}\]\,,\quad
{\H}_2=\/{Y_o\,t^{-2}}{(W-a^2_o)}\exp\[\int^z\!\!\/{\df
z'}{(W-a^2_o)}\],\quad {\H}_3=\/{Z_o}{t^2}\,,\nonumber
\end{align}
where $R_o$, $A_o$, $U_o$, $V_o$, $X_o$, $Y_o$ and $Z_o\in\R$; $R_o$
and  $Z_o\neq 0$, and $a^2_o={Z^2_o}/{R_o}$. The unknown function
$W$ of the symmetry variable $s=z$ is determined by solving the
integro-differential equation which corresponds to reduced form of
equation \Ref{s2} along the $z$-axis:
\begin{align}
W\/{dW}{dz}-W+&\/{A_o}{R_o}W^{-\g}\[2(2-\g)-\g\/{\df W}{\df z}\]
\exp\[2(2-\g)\int^z\!\/{\df
z'}{W}\]\label{int.85b}\\-&\/{(X^2_o+Y^2_o)}{ R_o}\/{W}
{(W-a^2_o)^3}\[\/{\df W}{\df z}-1\]\exp\[2\int^z\!\!\/{\df
z'}{(W-a^2_o)}\]=0\,.\nonumber
\end{align}
Equation \Ref{int.85b} is difficult to solve completely. However,
once again, with the ansatz \beq W=C_1 z+C_2,\;\quad\;
C_1,\;C_2\in\R,\;\quad\;C_1\neq 0\,,\nonumber\eeq we can transform
\Ref{int.85b} into an algebraic equation: \beq
(C_1-1)+\/{A_o}{R_o}(m+1)W^{m}
-\/{[X^2_o+Y^2_o]}{R_o}(C_1-1)\big(W-a^2_o\big)^{n}=0\,,
\nonumber\eeq where $m=[4-C_1+\g(C_1+2)]/C_1$ and $n=(2-3C_1)/C_1$.
Then:
\par
{\bf Case 1.)}\quad $C_1=1$,\quad $\g=4/3$
\begin{align}
\rho&=\/{R_o}{t^2(z+C_2)},\;\qquad\; p=\/{A_o}{t^4}\,,\label{II-85-a}\\
\displaystyle \v_1&=\/{(x-U_o)}{t}+\/{Z_o}{t}\/{X_o}{R_o},\qquad
\v_2=\/{(y-V_o)}{t}+\/{Z_o}{t}\/{Y_o}{R_o},\qquad
\v_3=\/{(z+C_2)}{t}\,,\cr {\H}_1&=\/{X_o}{t^2},\qquad
{\H}_2=\/{Y_o}{t^2},\qquad
{\H}_3=\displaystyle\/{Z_o}{t^2}\,,\nonumber
\end{align}
where $R_o\neq 0$, $A_o\in\R^{+}$; $C_2$, $U_o$, $V_o$, $X_o$,
$Y_o$, $Z_o\in\R$, $Z_o\neq 0$. To avoid singularities and to obtain
$\rho>0$, we must have  $t\neq 0$ and $z>-C_2$. Under these
conditions, this solution tends asymptotically to zero when
$t\rightarrow\infty$. It represents a compressible and irrotational
flow of a force-free fluid.
\par
{\bf Case 2.)}\quad $C_1=2/3$,\quad $\g=5/4$
\begin{align}
\rho&=\/{1}{t^2}\[2A_o+(X_o^2+Y_o^2)\]\left(\/{2}{3}z
+C_2\right)^{-1},\qquad p=\/{A_o}{t^4}\left(\/{2}{3}z
+C_2\right)\,,\label{85-b}\\
\displaystyle
\v_1&=\/{(x-U_o)}{t}+\/{a^2_o}{t}\/{X_o}{Z_o}\left(\/{2}{3}z
+C_2-a^2_o\right)^{1/2}\,,\cr
\v_2&=\/{(y-V_o)}{t}+\/{a^2_o}{t}\/{Y_o}{Z_o}\left(\/{2}{3}z
+C_2-a^2_o\right)^{1/2},\qquad \v_3=\/{1}{t}\left(\/{2}{3}z
+C_2\right)\,,\cr {\H}_1&=\/{X_o}{t^2}\left(\/{2}{3}z
+C_2-a^2_o\right)^{1/2},\qquad {\H}_2=\/{Y_o}{t^2}\left(\/{2}{3}z
+C_2-a^2_o\right)^{1/2},\qquad
{\H}_3=\displaystyle\/{Z_o}{t^2}\,,\nonumber
\end{align}
where $A_o\in\R^{+}$; $C_2$, $U_o$, $V_o$, $X_o$, $Y_o$, $Z_o\in\R$,
$Z_o\neq 0$. The parameter $a_o^2$ takes the form\beq
a^2_o={Z_o^2}\big/{\[2A_o+(X_o^2+Y_o^2)\]}\,,\nonumber\eeq which can
be interpreted as the square of the Alfv\'en velocity. The solution
\Ref{85-b} is real and non-singular if $z> 3(a^2_o-C_2)/2$ and
$t\neq 0$, and tends asymptotically to zero when
$t\rightarrow\infty$. This solution describes the propagation of a
compressional Alfv\'en wave in an ideal fluid
\cite{pic},\cite{swanson}. This can be illustrated with the
expression of the Lorentz force:
\beq\Fm=\/{Z_o}{3t^4}\left(\/{2}{3}z
+C_2-a^2_o\right)^{-1/2}\big[X_o\,\ve_1+Y_o\,\ve_2\big]-\/{(X^2_o+Y^2_o)}{3t^4}\,\ve_3\,.\eeq
The first term on the right-hand side of this equation represents
the action of a tension force along the magnetic field lines that
acts only in the $x$-$y$ plane (they twist relatively to one
another, but do not compress), while the second term corresponds to
the gradient of magnetic pressure, acting perpendicular to the
$x$-$y$ plane, which causes compressions and expansions of the
distance between the lines of force without changing their
direction. So, the solution \Ref{85-b} describes a basic oscillation
between kinetic fluid energy (fluid inertia) and compressional
(field line pressure) plus line bending (field line tension)
magnetic energy. The vorticity of the flow is given by
\beq\vw=\/{a^2_o}{3Z_o}t^{-1}\left(\/{2}{3}z
+C_2-a^2_o\right)^{-1/2}\Big(-Y_o,X_o,0\,\Big)\,.\eeq However, by
virtue of Kelvin's theorem, the circulation of the fluid is not
conserved because of the presence of tension force originating from
the Lorentz force: the term on the right-hand of the equation
\Ref{vort} is nonzero.


\section{Concluding remarks}

We summarize the main results achieved in this paper. Using the
subgroup structure of the symmetry group of the MHD equations
\Ref{s1}-\Ref{s5} to compute G-invariant solutions (GIS) of this
system of PDEs, we have considered three-dimensional subgroups
${\cal G}$, where ${\rm{codim}}[{\cal G}]=1$ in the space of
independent variables $E$, in order to obtain reduced systems
strictly composed of ODEs. The Lie algebras of these groups are
representatives of the conjugacy classes of the Galilean-similitude
(GS) algebra that have been calculated previously by Grundland and
Lalague \cite{amg0}. By means of a procedure of symmetry reduction
of Eqs \Ref{s1}--\Ref{s5}, we have found some novel (at the best of
our knowledge) exact solutions which demonstrate the efficiently of
the SRM. Such solutions have been discussed by the point of view of
their possible meaning.

In Section 4, we present solutions which describe a nonstationary
and compressible flow in the presence of a magnetic field
$\vH=({\H}_1,{\H}_2,0)$. In particular, the Lorentz force (except
for solution \Ref{3-911a} where $\Fm=\boldsymbol{0}$) acts
perpendicularly to $\vH$, causing compressions and expansions of the
lines of force without changing their direction, as does a
magnetoacoustic fast wave $\rm{F}$ which propagates perpendicularly
to $\vH$ in an ideal fluid. There is no tension force, so the
circulation of the fluid is conserved, which is also meaning that
the vortex lines are frozen into the fluid.

In Section 5, we obtained several types of solutions where
$\vH=({\H}_1,{\H}_2,{\H_3})$. For the algebras $G_4$, $G_5$ and
$G_6$ we find stationary solutions for which the fluid is
incompressible and its flow is along the axis of symmetry in the
context of slab, cylindrically and spirally geometry, respectively.
Solution $G_7$ describes the propagation of a nonstationary Alfv\'en
wave in an incompressible fluid: $\Fm$ is purely a tension force and
the circulation of the fluid is not conserved. Solution $G_8$
represents a nonstationary and compressive flow of a force-free
fluid in $(3+1)$ dimensions. Solution $G_{9}$ describes the
propagation of a double magnetoacoustic wave $\rm{F}\rm{F}$ in a
nonstationary and compressible fluid: $\Fm$ causes compressions and
expansions of the lines of force along both the $x$- and $z$-axes.
There is no tension force, so the circulation of the fluid is
conserved. Finally for the algebra $G_{10}$, we obtain solutions
\Ref{II-85-a} and \Ref{85-b}. The first represents a compressible
and irrotational flow of a force-free fluid. The second one
corresponds to the propagation of a ``compressional Alfv\'en wave''
in an ideal fluid. Namely, it combines the action of tension force
along the lines of force and the compressions and expansions of
those resulting from magnetic pressure force. Obviously, the
circulation of the fluid is not conserved.

It is suitable to recall than an approach of group-theoretical type
to the equations for the fluid dynamics, especially to the
isentropic compressible fluid model and to the MHD equations, is not
new and has been carried out by many authors (see, for example,
\cite{fuchs}, \cite{amg1}-\cite{popo}). On the other hand, a full
classification of Lie subgroups of the symmetry group of MHD
equations in $(3+1)$ dimensions was never been obtained before
\cite{amg0}, and therefore allows us for a systematic approach to
the task of constructing both invariant and partially invariant
solutions of Eqs \Ref{s1}--\Ref{s5}. The objective of future study
is to take full advantage of some recently developed alternative
versions of SRM, namely the method of partially invariant solutions
(PIS) and the weak transversality method (WTM).

The notion of PIS was introduced by Ovsiannikov \cite{ovs} and it
relates to the case when only a part of the graph $\Gamma_{f}$ of
the solution $\u=f(\x)$ is G-invariant. This means that a solution
$\u=f(\x)$ is a partially invariant solution if the number \beq
\delta={\rm{dim}}\,({\cal G}\Gamma_{f})-p\,,\eeq (where ${\cal
G}\Gamma_{f}$ is the orbit of the graph $\Gamma_{f}$), called the
defect structure of the function $f(\x)$ with respect to the group
${\cal G}$, satisfies the condition \beq
0<\delta<{\rm{min}}\,(r,q)\,,\eeq where $r$ is the dimension of the
orbits of ${\cal G}$ in $E\times U$. The construction of this type
of solution leads, through the reduction of the original system
$\Delta$, to two differential systems of equations: $\Delta^{1}$ for
$\delta$ dependent and $p$ independent variables and $\Delta/{\cal
G}$ for $(q-\delta)$ dependent and $(p+\delta-r)$ independent
variables. There is no longer a one-to-one correspondence between
PIS of $\Delta$ and the solutions of $\Delta/{\cal G}$. For one
solution of $\Delta/{\cal G}$ we have a family of solutions of
$\Delta$. In other words, this approach generates more solutions,
covering much larger classes of initial and boundary conditions than
the classical method. Nevertheless, there have been very few
examples of its application. Construction of PIS requires a much
more complex procedure than the ``classical'' SRM, but Grundland and
Lalague \cite{amg2} developed an effective algorithmic tool for this
purpose.

The future study of MHD equations will focus on the construction of
PIS invariant under four-dimensional subgroups (and some
three-dimensional subgroups as mentioned above) with the defect
structure $\delta=1$. Under these assumptions the original equations
can be reduced to the systems $\Delta/{\cal G}$ of ODEs, and one PDE
denoted by $\Delta^1$ for one unknown function.  It could be mentioned also that the procedure cited above produces many new reducible solutions which would be,
paradoxically, more difficult to obtain by the SRM, since they are
related to lower dimensional subgroups than ${\cal G}$, leading to
the reductions of $\Delta$ to PDEs rather than ODEs.

The second component of a future project will involve the recently
WTM \cite{amg3}. This new method is based on a ``group invariant
solutions without transversality'' approach developed by Anderson
{\it et al.} \cite{anderson} which was designed to overcome the
limitation of the SRM resulting from the transversality requirement,
and extends the applicability of this classical method. The notion
of transversality refers to the condition imposed on the vector
fields generating subalgebras of a given system. Consider a
$r$-dimensional subgroup ${\cal G}_o\subset{\cal G}$, or its
corresponding subalgebra ${\cal L}_o\subset{\cal L}$, generated by
vector fields \beq\hat
X_a=\xi_a^\mu\,(\x,\u)\partial_{x^\mu}+\phi_a^j\,(\x,\u)\partial_{u^j}\,,
\quad\qquad a=1,\ld,r\eeq with the matrix of characteristics of the
vector fields $\hat X_a$ spanning the subalgebra ${\cal L}_o$ \beq
Q_a^{j}(\x,\u^{(1)})=\left\{\phi_a^j\,(\x,\u)-\xi_a^\mu\,
(\x,\u)\Dp{u^j}{x^\mu}\right\}\,,\qquad a=1,\ld,r\,,\quad
j=1,\ld,q\,.\label{matrix}\eeq We require that the subgroup ${\cal
G}_o$ acts regularly and transversally on the manifold ${\cal
M}=E\times U$. It means that for each point $(\x,\u)\in{\cal M}$ the
relation \beq{\rm rank}\big\{\xi^{\mu}_a(\x,\u)\big\}={\rm
rank}\big\{\xi^{\mu}_a(\x,\u),\phi^{j}_a(\x,\u)\big\}\label{rank}\eeq
holds. When this condition, which we call the ``strong
transversality'' condition, is not satisfied for a given subalgebra,
in principle the classical SRM cannot be applied, \ie\, the rank of
the Jacobian matrix \Ref{step3} is not maximal. It still possible,
however, that there exists a certain domain ${\cal M}_o\subset {\cal
M}$ for which this condition is fulfilled. To distinguish such
cases, Grundland {\it et al.} \cite{amg3} have introduced the notion
of ``weak transversality'' and have shown that subalgebras with this
property can still be used to construct invariant and partially
invariant solutions using a specific algorithm developed for this
purpose. In order to do it, we determine the conditions on $\u =
f(\x)$ under which rank requirement \Ref{rank} is satisfied and
solve these conditions to obtain the general form of these
functions. Next we substitute the obtained expressions into the
matrix of characteristics \Ref{matrix} and require that the
condition ${\rm rank}\,Q=0$ is satisfied. This provides further
constraints on the functions $f(\x)$. Finally, we solve the
overdetermined system obtained from the system under investigation
subjected to the constraints coming from requirements of weak
transversality. If solutions of this overdetermined system exist
then invariant solutions can be constructed explicitly.

{}
\label{lastpage}
\end{document}